# A Feature-based Classification Technique for Answering Multi-choice World History Questions: FRDC_QA at NTCIR-11 QA-Lab Task

Shuangyong Song, Yao Meng, Zhongguang Zheng, Jun Sun
Internet Application Laboratory, Fujitsu R&D Center CO., LTD
No.56 Dong Si Huan Zhong Rd., Chaoyang District,
Beijing 100025, China
{shuangyong.song, mengyao, zhengzhg, sunjun}@cn.fujitsu.com

## ABSTRACT
Our FRDC_QA team participated in the QA-Lab English subtask of the NTCIR-11. In this paper, we describe our system for solving real-world university entrance exam questions, which are related to world history. Wikipedia is used as the main external resource for our system. Since problems with choosing right/wrong sentence from multiple sentence choices account for about two-thirds of the total, we individually design a classification based model for solving this type of questions. For other types of questions, we also design some simple methods.

## Team Name
FRDC_QA

## Subtask
QA-Lab English Subtask

## 1. INTRODUCTION

Question Answering (QA) is a specialized area in Information Retrieval. QA systems are concerned with providing relevant answers in response to questions proposed in natural language. QA is therefore composed of three distinct modules: question classification, information retrieval, and answer extraction, each of which has a core component beside other supplementary components [7]. Question classification plays an essential role in QA systems by classifying the submitted question according to its type.

In particular, solving real-world school exam questions is an important and useful application of QA systems, and some research has been done on this task [1, 8-10]. NTCIR-11 QA-Lab task aims to provide a module-based platform for system performance evaluations and comparisons for solving real-world university entrance exam questions, which are selected from The National Center Test for University Admissions and from secondary exams at 4 universities in Japan.

FRDC_QA take part in the English subtask. We design a system and the details of it are given as follows: In section 2, we introduce the external resource and some convenient storage ways. The framework for questions with multiple sentence choices is proposed in section 3. In section 4, the frameworks for other types of questions are described. The evaluation results of our system on world history exam B in 2007 Japan University Admissions are given in section 5. Finally we make a conclusion and discuss our plans for future work in section 6.

## 2. HASH MAP & LUCENE INDEXES OF EXTERNAL RESOURCE

### 2.1 External Resource

We utilize Wikipedia as external resource for our QA-Lab task. Wikipedia is a well-known free content, multilingual encyclopedia written collaboratively by contributors around the world [6]. In this paper, for the English subtask, we download the Wikipedia dataset with the version of 'enwiki dump progress on 20140502' from Wikimedia Downloads [1]. Downloaded dataset contains 'enwiki-20140502-all-titles' as the list of all the Wiki-items, and 'enwiki-20140502-pages-articles' with articles of all the Wiki-items. All those data will be processed to be more formal and then be stored in hash map or Lucene[2] indexes, for realizing convenient and quick search in our QA system. The details are given below.

### 2.2 Hash Map of Item Title

For quickly checking if a word or word group is a Wikipedia item, we put all Wikipedia item titles into a hash map. The title list dataset contains 32,877,103 titles of Wikipedia items in total, and we convert all characters of them to be lowercase. Word or word groups will also be converted to be characters in lowercase when they are checked, for realizing an exact matching. When we detect items contained in a sentence, we adopt a Maximum Matching Method. For example, for a sentence with $N$ words, we first check if this whole sentence is a Wiki item, and then check all sub-sentences with $N$-1 continuous words, then sub-sentences with length of $N$-2, and so on. In particular, if a detected item consist of another detected item, the latter one will be removed and the longer one will be reserved.

### 2.3 Lucene Index of Item Page

Each item has its related Wiki article, describing the details of this item. We put the title and page content into a Lucene index file as two word string fields, and then we can easily get the description of a Wiki item with simple Lucene search. This index file is used to search the relationship between items, since two related items will show in the Wiki article of each other, vice versa.

---

[1] http://download.wikipedia.com/enwiki/20140502
[2] http://lucene.apache.org

## 2.4 Lucene Index of Item Redirection

Different Wiki items may have same meanings, such as 'AccessibleComputing' and 'Computer accessibility'. For those items with same meanings, Wikipedia utilize 'redirection' tag to link one of them to another. Therefore, just one of them has a Wiki article with detailed description, and the Wiki article of another item just contains one sentence with redirection declaration. Take the 'AccessibleComputing' and 'Computer accessibility' for example, the Wiki article of 'Computer accessibility' contains detailed description of this item, but the Wiki article of 'AccessibleComputing' just contains one sentence "Redirect page → Computer accessibility". We put those 'AccessibleComputing' like redirected Wiki item titles into a Lucene index file as one word string field, and take the related item titles as another word string field, then we can easily search the real description of those redirected Wiki items.

## 2.5 Lucene Index of Item Time

For answering some questions about items' occurrence time, we extract the time information of each item from the Wiki articles and put those information into a Lucene index. There are two different types of time information we should extract, one is the exact time of this item, and another is the period of this item. Such as the time of 'Independence Day (United States)' is 'The Fourth of July', and the period of 'French and Indian War' is '1754–1763'. For the period type, we respectively record the front part as 'start time' and latter part as 'end time' since lots of questions may ask them separately.

## 3. FRAMEWORK FOR QUESTIONS WITH MULTIPLE SENTENCE CHOICES

## 3.1 Brief Description of Framework

We first introduce the type of questions with multiple sentence choices, and an example is given below (italic characters):

*Background text: ... ... (8) In India in the latter half of the 19th century, a large-scale rebellion against colonial rule took place; one of the things that triggered this was the fact that Muslim soldiers revolted due to a rumor that pork fat had been used on the cartridges in their guns. ... ...*

*Question: From 1-4 below, choose the most appropriate sentence concerning the underlined portion (8).*

*Choices:*
1. *This rebellion is also called the Sipahi (Sepoy) Mutiny.*
2. *The Ever Victorious Army was actively involved in suppressing this rebellion.*
3. *The Ever Victorious Army was actively involved in suppressing this rebellion.*
4. *After this rebellion, Queen Victoria also became the empress of the Mughal Empire.*

We take this type of questions as accuracy probability ranking problem for all the choices, and we utilize classification models to handle this problem. By separating right choices and wrong choices in the training dataset, binary classification models can be trained. Seven features, such as semantic relationship between background text and choices, semantic relationship between question and choices, etc., are extracted for training classification models, and eleven classifiers are selected to calculate the accuracy probability of each choice. Then the results of those classifiers are combined together to get the final ranking of choices. Details of this framework are given in following subsections.

## 3.2 Features

We extracted seven features of each choice in total for classifiers, and the details of them are given below:

1) Internal Item Relativity:

We first detect all items contained in a choice sentence with the Maximum Matching Method described in section 2.2, and then detect relationships among those items. The method for judging if or not two items are related is detecting if an item shows in the Wiki article of another item. For an choice sentence consisting of $N$ items, we can get $N(N-1)/2$ 'item couple', and each related 'item couple' will contribute 1 point to this feature. Therefore, the value of this feature 'Internal Item Relativity' will be from 0 to $N(N-1)/2$.

2) Item Relativity between Text and Choice:

All the items contained in the 'text portion' and choice sentence need to be detected first, and relationships between Text items and Choice items will be detected. For a text sentence consisting of $M$ items and a choice sentence consisting of $N$ items, we can get $M*N$ 'item couple', and each related 'item couple' will contribute 1 point to this feature. Therefore, the value of this feature 'Item Relativity between Text and Choice' will be from 0 to $M*N$.

3) Item Relativity between Question and Choice:

All the items contained in the question sentence and choice sentence need to be detected first, and relationships between Question items and Choice items will be detected. For a question sentence consisting of $Q$ items and a choice sentence consisting of $N$ items, we can get $Q*N$ 'item couple', and each related 'item couple' will contribute 1 point to this feature. Therefore, the value of this feature 'Item Relativity between Question and Choice' will be from 0 to $Q*N$.

4) Minimum Distance with Negative Sentences:

We assume that one choice 'more similar with a negative sentence, more likely to be a wrong answer'. For getting this feature, we firstly need to exact all the negative sentences in Wiki articles, which contains 'is not', 'are not', 'did not' or other negative expressions. After removing stop words in choice sentences and those negative sentences, all of them can be represented as word vectors. Distance between two word vectors $V_1$ and $V_2$ is calculated with the formula below:

$$D(V_1,V_2) = \sum_{i=1}^{L(V_1 \cup V_2)} (w_{1i} - w_{2i})^2 \qquad (1)$$

in which, the $D(V_1,V_2)$ means distance between vector $V_1$ and vector $V_2$, and the $V_1 \cup V_2$ means the union of $V_1$ and $V_2$, and the $L(V_1 \cup V_2)$ means the length of $V_1 \cup V_2$, and the $w_{1i}$ means the value of $V_1$ on the $i^{th}$ dimension of $V_1 \cup V_2$, and the $w_{2i}$ means the value of $V_2$ on the $i^{th}$ dimension of $V_1 \cup V_2$. This formula is modified from the Euclidean Distance [2], without firstly creating the vector of words in all Wiki articles and choices, which is very time consuming and with low robustness.

5) Number of Related Wiki Articles:

With the 'Lucene Index of Item Page' described in section 2.3, we take a choice as a query and search all the possible related Wiki articles from this index file. The number of returned Wiki articles will be taken as the value of this feature.

6) Similarity with Top 1 Related Wiki Article:

The search method is same as the above feature, but the value of this feature is the value of the semantic similarity between the choice sentence and the top 1 returned Wiki article, which is very easy to get with a ready-made function in Lucene system.

7) Similarity with Top 3 Related Wiki Articles:

The search method is same as the above feature, but the value of this feature is the average value of the semantic similarity between the choice sentence and the top 3 returned Wiki articles.

### 3.3 Classifiers

In our system, we in total utilize eleven classifiers to training different classification models respectively. Simple description of them are given below:

*Random Forest*: Random forest is an ensemble learning method for classification that operate by constructing a multitude of decision trees at training time and outputting the class that is the mode of the classes output by individual trees [3].

*LogitBoost*: LogitBoost is a boosting algorithm that casts the AdaBoost algorithm into a statistical framework. Specifically, if one considers AdaBoost as a generalized additive model and then applies the cost functional of logistic regression, one can derive the LogitBoost algorithm [4].

*Logistic Model Trees*: Logistic model tree (LMT) is a classification model with an associated supervised training algorithm that combines logistic regression (LR) and decision tree learning [11].

*AdaBoost M1*: AdaBoost M1 is an improved version of traditional AdaBoost algorithm, which can be used to classify both binary and polynominal label with numerical, binominal and polynominal (and weighted) attributes [12].

*Bagging*: Bagging, also called bootstrap aggregating, is a machine learning ensemble meta-algorithm designed to improve the stability and accuracy of machine learning algorithms used in statistical classification and regression. It also reduces variance and helps to avoid over-fitting [13].

*MultiBoostAB*: MultiBoosting is an extension to the highly successful AdaBoost technique for forming decision committees. MultiBoosting can be viewed as combining AdaBoost with wagging. It is able to harness both AdaBoost's high bias and variance reduction with wagging's superior variance reduction [14].

*Locally Weighted Learning*: Locally weighted learning uses an instance-based algorithm to assign instance weights which are then used by a specified Weighted Instances Handler. Can do classification (e.g. using naive Bayes) or regression (e.g. using linear regression) [15].

*Logistic Regression*: Logistic regression is a probabilistic statistical classification model, which can be used to predict a binary response from a binary predictor, used for predicting the outcome of a categorical dependent variable based on one or more predictor variables. [16].

*Simple Naïve Bayes*: Naive Bayes classifier, in which the numeric attributes are modelled by a normal distribution [17].

*Naïve Bayes*: An improved Naive Bayes classifier using estimator classes. Numeric estimator precision values are chosen based on analysis of the training data [18].

*Updateable Naïve Bayes*: An updateable version of Naïve Bayes model. This classifier will use a default precision of 0.1 for numeric attributes when build Classifier is called with zero training instances [18].

All of the classification model training procedures are realized with WEKA [5], which is a collection of machine learning algorithms for data mining tasks and contains tools for data pre-processing, classification, regression, clustering, association rules, and visualization.

### 3.4 Choice Selection

Each classifier can get an accuracy probability for each choice, and the average value of the accuracy probability from all classifiers will be taken as the final accuracy probability of a choice. Then, if the question is asking us to choose the right choice with the keywords 'correct', 'correctly' or 'appropriate', we choose the choice with highest accuracy probability as the final answer. If the question is asking us to choose the wrong choice with the keywords 'incorrect', 'incorrectly' or 'mistake', we choose the choice with lowest accuracy probability as the final answer.

## 4. FRAMEWORKS FOR OTHER TYPES OF QUESTIONS

In this section, we give some description of our frameworks for problems besides the type with multiple sentence choices, such as questions with chronological sequence choices, questions with term choices, etc.

### 4.1 Framework for Questions with Chronological Sequence Choices (without images)

We first give an example of this type of question (italic characters):

*Background text: ... ... (7) A Cold War began between the US and the Soviet Union, and the world faced another serious conflict. In relation to this ... ...*

*Question: In regard to the underlined portion (7), from (1)-(4) below, choose the correct chronological sequence of events relating to the Cold War.*

*Choices:*
1. *Warsaw Treaty Organization formed - Berlin blockade - Cuban missile crisis - Japan-US Security Treaty signed (1951)*
2. *Berlin blockade - Japan-US Security Treaty signed (1951) - Warsaw Treaty Organization formed - Cuban missile crisis*
3. *Japan-US Security Treaty signed (1951) - Cuban missile crisis - Berlin blockade - Warsaw Treaty Organization formed*
4. *Berlin blockade - Warsaw Treaty Organization formed - Japan-US Security Treaty signed (1951) - Cuban missile crisis*

For this type of questions, we utilize the 'Lucene Index of Item Time' to search timestamp of each event in the choices, and rank them with the chronological order, then we can choose the right answer according to this order easily.

## 4.2 Framework for Questions with Term Choices (without images)

An example of this type of question is given below (italic characters):

*Background text: ... ... (1) Nomadic tribes on horseback emerged on the Eurasian continent. Their elusive character became a major threat to sedentary agricultural societies, so troops mounted on horseback were organized to counteract them. Rulers who sought good horses also emerged, such as ... ...*

*Question: In regard to the underlined portion (1), from 1-4 below, choose the one name that correctly describes the nomadic tribe on horseback that came to prominence in the 6th century and built up a nation.*

*Choices:*
1. *Scythians*
2. *Göktürks*
3. *Yuezhi*
4. *Xiongnu*

We detect items contained in the background text and the question with the Maximum Matching Method, then calculate the relativity between those items and the choice item, with using the same method described in section 3.2. Finally, the choice with highest relativity with the background text and the question will be chosen as the final answer.

## 4.3 Framework for Questions with Judging True or False Sentences (without images)

An example of this type of question is given below (italic characters):

*Background text: ... ... (3) founder of the kingdom - is believed to be the Chumo who appears in the "Book of Wei (Weishu)", which is a record of the Northern Wei dynasty... ...*

*Question: In regard to the underlined portion (3), from 1-4 below, choose the correct combination of "correct" and "incorrect" in regard to the following sentences taa and b concerning the historic founder of the kingdom.*

*Question text: a Liu Bang defeated Xiang Yu and made Chang'an the capital. b Yelü Dashi built the Kara-Khitan Khanate.*

*Choices:*
1. *a - Correct b - Correct*
2. *a - Correct b - Incorrect*
3. *a - Incorrect b - Correct*
4. *a - Incorrect b - Incorrect*

We use the same training data with same features as described in section 3.2 to train Support Vector Machine classification model (SVM) to handle this type of questions by directly output the 'true of false' result of each choice instead of the accuracy probability. Then we can easily choose the right choice according to the output of the SVM model.

## 4.4 Framework for Other types of Questions

We choose the final answer with the random selection method for other types of questions, which usually need image analysis technology. In particular, we set a specified random seed to keep the stability of the results given by our system.

## 5. EVALUATION RESULTS

Table 1 gives the evaluation results of our system on the phase 1 contest data - world history exam B in 2007 Japan University Admissions.

**Table 1. Evaluation results of our system in phase 1**

| Type of questions | Number of correct answer / Total number | Score of correct answer / Total score |
|---|---|---|
| Questions with multiple sentence choices | 10/23 | 28/62 |
| Questions with chronological sequence choices (without images) | 0/0 | 0/0 |
| Questions with term choices (without images) | 3/7 | 9/20 |
| Questions with judging true or false sentences (without images) | 0/0 | 0/0 |
| Other types of questions | 0/7 | 0/18 |
| Total | 13/36 | 37/100 |

For types 'Questions with multiple sentence choices' and 'Questions with term choices (without images)', we achieve a precision of about 45% on both 'Number of correct answer' and 'Score of correct answer', which shows the much better effectiveness than random method, since we think the precision of random method should be 25% on four-choice questions. However, the real result of random method on 'other types of questions' is not as good as our thought. We got wrong answers on all the seven 'other types of questions' with the random method, which makes our total result getting a precision of 37%, far below the 45%.

## 6. CONCLUSIONS AND FUTURE WORK

In our work of NTCIR-11 QA-Lab task, we design a system for solving real-world university entrance exam questions, which are related to world history. We utilize Wikipedia as the main external resource for our system, since nearly all of world history knowledge can be found in Wikipedia. In addition, we design different solution frameworks for different types of questions, such as questions with multiple sentence choices, questions with temporal term choices, questions with non-temporal term choices, etc. Although our system performs much better than random methods, it is still far from meeting actual demand. Several attempts can be tried to improve the system performance in our future work, e.g., (1) more useful external resources can be utilized, such as query results from Google like search engines, electronic history books, etc. (2) more reasonable and intelligent combination way for different classification models should be tried; (3) different writing styles for timestamps, locations and personal names should be considered. Furthermore, a unified domain insensitive system for choosing wrong/right answer from multiple sentence choice will be a trial in our future work.

## 7. ACKNOWLEDGMENTS

We sincerely thank our colleagues in Fujitsu Laboratories Ltd., Takuya Makino, Hiroko Suzuki, Tomoya Iwakura and Tetsuro Takahashi, for their help on providing us external training dataset and valuable discussions with us about feature extraction and machine learning methods.

## 8. REFERENCES

[1] Kano, Y. 2014. Solving History Problems of the National Center Test for University Admissions. In *Proceedings of*

the 28th Annual Conference of the Japanese Society for Artificial Intelligence.

[2] Song, S., Li, Q. and Bao, H. 2012. Detecting Dynamic Association among Twitter Topics. In *Proceedings of the 21st International World Wide Web Conference* (Lyon, France, April 16-20, 2012), pages 605-606.

[3] Breiman, L. 2001. Random forests. *Machine Learning*. 45(1): 5–32.

[4] Friedman, J., Hastie, T. and Tibshirani, R. 2000. Additive logistic regression: a statistical view of boosting. *Annals of Statistics*. 28(2): 337–407.

[5] Hall, M., Frank, E., Holmes, G., Pfahringer, B., Reutemann, P., Witten, I. H. 2009. The WEKA Data Mining Software: An Update. *SIGKDD Explorations*. Volume 11, Issue 1.

[6] Bhole, A., Fortuna, B., Grobelnik, M. and Mladenić, D. 2007. Extracting Named Entities and Relating Them over Time Based on Wikipedia. *Informatica (Slovenia)*. 31(4): 463-468.

[7] Allam, A. M. N., Haggag, M. H. 2012. The Question Answering Systems: A Survey. *International Journal of Research and Reviews in Information Sciences*. Vol. 2, No. 3, September 2012.

[8] Martin R, F., Sibyl, H., Veronika, K. 2005. Answering multiple-choice questions in high-stakes medical examinations. *Medical Education*, Vol. 39, No. 9, September 2005, pp. 890-894.

[9] Awadallah, R., Rauber, A. 2006. Web-Based multiple choice question answering for english and arabic questions, In *Proceeding of the 28th European conference on Advances in Information Retrieval*, pp. 515-518.

[10] Sorger, B., Dahmen, B., Reithler, J., Gosseries, O., Maudoux, A., Laureys, S., Goebel, R. 2009. Another kind of 'BOLD Response': answering multiple-choice questions via online decoded single-trial brain signals. *Progress in Brain Research*, Vol. 177, 2009, pp. 275–292.

[11] Landwehr, N., Hall, M. A., Frank, E. 2003. Logistic Model Trees. In *Proceeding of the 14th European Conference on Machine Learning*. pp. 241-252.

[12] Eibl, G., Pfeiffer, K. P. 2002. How to Make AdaBoost.M1 Work for Weak Base Classifiers by Changing Only One Line of the Code. In *Proceeding of the European Conference on Machine Learning and Principles and Practice of Knowledge Discovery in Databases*. pp. 72-83.

[13] Breiman, L. 1996. Bagging predictors. *Machine Learning*, vol. 24, no. 2, pp. 123-140.

[14] Webb, G. I. 2000. MultiBoosting: A Technique for Combining Boosting and Wagging. *Machine Learning*. Vol.40, no. 2, pp. 159-196.

[15] Atkeson, C. G., Moore, A. W., Schaal, S. 1997. Locally Weighted Learning. *Artificial Intelligence Review*, vol. 11, no. 1, pp. 11-73.

[16] Bishop, C. M., Nasrabadi, N. M. 2007. Pattern Recognition and Machine Learning. Jour*nal of Electronic Imaging*, vol. 16, no. 4.

[17] Duda R., Hart, P. 1973. Pattern Classification and Scene Analysis. Wiley, New York.

[18] John, G. H., Langley, P. 1995. Estimating Continuous Distributions in Bayesian Classifiers. In *Proceedings of the Eleventh Conference on Uncertainty in Artificial Intelligence*. pp. 338-345.